\documentclass[12pt]{iopart}
\usepackage{graphicx} 
\newcommand{\jpsi}{\rm J/$\psi$}

\begin{document}

\title[\jpsi\ production in pp collisions at $\sqrt{s}$=2.76 and 7 TeV with ALICE]{Measurement of $\rm J/\psi$
production in pp collisions at $\sqrt{s}$=2.76 and 7 TeV with ALICE}

\author{Roberta Arnaldi for the ALICE Collaboration}

\address{INFN Sez. Torino, Via P. Giuria 1, I-10125, Torino, Italy}
\ead{arnaldi@to.infn.it}
\begin{abstract}
We present results from the ALICE experiment on the  
inclusive \jpsi\ production in pp collisions at $\sqrt{s}$=2.76 and 7~TeV. 
The integrated and differential cross sections are evaluated down to 
$p_{\rm T}$=0 in two rapidity ranges, $|y|<0.9$ and $2.5<y<4$, in the 
dielectron and dimuon decay channel respectively. 
The measurement at $\sqrt{s}$=2.76~TeV, the same energy as 
Pb-Pb collisions, provides a crucial reference for 
the study of hot nuclear matter effects on \jpsi\ production.  
The \jpsi\ yield in pp collisions at $\sqrt{s}$=7~TeV has  also been studied as a
function of the charged particle multiplicity and first results are presented.
\end{abstract}
\vskip -0.8cm
\vskip 0.5cm

In 2010, the Large Hadron Collider (LHC) provided, for the first time, pp 
collisions at $\sqrt{s}$=7~TeV, followed in 2011, by pp collisions at a lower
energy, $\sqrt{s}$=2.76~TeV. 
Quarkonium production study in these interactions has a twofold interest.
On one hand, this new energy regime might allow further insight in the 
understanding of the quarkonium production mechanisms. 
In fact, theoretical models have not been able, up to now, to consistently interpret, 
at RHIC and Tevatron energies, all aspects of quarkonium production, as the 
\jpsi\ transverse momentum ($p_{\rm T}$) distribution and polarization.
On the other hand, pp collisions represent a 
reference to understand the quarkonium behaviour in the hot and dense state of
QCD (or hadronic) matter formed in heavy ion collisions. In particular, pp
interactions at $\sqrt{s}$=2.76~TeV, turn out to be a crucial baseline, being
collected at the same $\sqrt{s_{NN}}$ of the Pb-Pb collisions.

The ALICE experiment consists of a central barrel ($|\eta|<0.9$) and a forward muon 
spectrometer ($-4<\eta<-2.5$). 
Central barrel detectors are located inside a large solenoidal magnet and allow 
the tracking of particles down to $p_{\rm T}$ of about 
100 MeV/$c$. 
The barrel detectors, used in this analysis, 
are the Inner Tracking System (ITS), for the vertex identification and track 
reconstruction close to the interaction point, and the Time Projection Chamber 
(TPC) for tracking and electron identification.
Muons with a momentum larger than 4~GeV/$c$ are detected in the muon
spectrometer.
It consists of a front absorber, followed by a dipole magnet, coupled to a set 
of tracking and triggering chambers. 
An iron wall, placed upstream the trigger chambers, absorbs secondary hadrons
escaping the front absorber. Finally, the forward VZERO detector, made of two 
scintillator arrays covering the range 2.8$<\eta<$ 5.1 and -3.7$<\eta<$-1.7, is used for triggering purposes. For more details on the
apparatus, see~\cite{Aam08}.
Quarkonia are measured in the dielectron decay channel, at midrapidity, 
and in the dimuon one, at forward rapidity. 
The transverse momentum acceptance goes down to 
$p_{\rm T}$=0, in both $y$ regions, providing, at midrapidity, a
unique coverage at LHC.

In this proceeding, we present results on the inclusive \jpsi\ production, while
the analysis of the 2011 
higher statistics sample will also allow us to measure the prompt \jpsi\ fraction at
midrapidity.
Final results based on a fraction of the 2010 pp data at 
$\sqrt{s}$=7~TeV and preliminary results from the pp data collected in a three-day long run 
in 2011 at $\sqrt{s}$=2.76~TeV are discussed.
The data sample corresponds to events collected with the minimum bias 
trigger, defined as the logical OR between the requirement of at 
least one hit in the ITS pixel layers (SPD), and a signal in one 
of the two VZERO detectors. For the muon analysis, the minimum
bias trigger is required to be in coincidence with a signal in the muon trigger
chambers. 

For the J/$\psi\rightarrow\mu^+\mu^-$ analysis, events are 
selected asking for at least one interaction vertex reconstructed by the SPD and applying quality cuts on the muon tracks. 
Furthermore, to reject hadrons produced in the front absorber, at least 
one of the two muons, forming the dimuon, should also give a signal in the trigger chambers. Requiring both muons to
satisfy the trigger condition does not significantly improve the 
purity of the \jpsi\ sample, in spite of a significant loss ($\approx$
20\%) in statistics.
A fit to the dimuon invariant mass spectrum, in the region $1.5<m_{\mu\mu}<5$~GeV/$c^{2}$ allows to extract the \jpsi\ yield. The background
is described by a double exponential function, while the signals, \jpsi\ and
$\psi(2S)$, are fitted with Crystal Ball functions\cite{Gai82}. 
At $\sqrt{s}$=7~TeV, the number of \jpsi\, corresponding to an integrated 
luminosity ($L_{\rm int}$) of 15.6 nb$^{-1}$, is 
$1924\pm77$.

For the J/$\psi\rightarrow\e^+\e^-$ analysis, electrons are identified through the
specific energy loss in the TPC, applying a $\pm$3$\sigma$ inclusion cut and a 
$\pm$3.5 ($\pm$3)$\sigma$ exclusion cut for pions (protons).
The \jpsi\ yield is evaluated from a bin 
counting technique, after subtracting the like-sign 
background to the invariant mass spectrum. The total number of \jpsi, 
corresponding to $L_{\rm int}=3.9$ nb$^{-1}$, is $N_{\rm J/\psi}= 249\pm 27$.
A Monte Carlo simulation, based on realistic \jpsi\ $p_{\rm T}$ and $y$ 
distributions, is used to estimate the acceptance of the apparatus and  
the reconstruction and trigger efficiencies ($A \times \epsilon$).
At $\sqrt{s}$=7~TeV, for the J/$\psi\rightarrow\e^+\e^-$ analysis 
$A \times \epsilon$
is 10\%, while for the J/$\psi\rightarrow\mu^+\mu^-$ analysis it is
 33\%.
The inclusive \jpsi\ production cross section is obtained dividing the 
number of \jpsi\ by $A \times \epsilon$ and by the 
integrated luminosity.
Results at $\sqrt{s}$=7~TeV are: 
$\sigma_{\rm J/\psi}(2.5<y<4)$=6.31$\pm$0.25~(stat.)$\pm$0.80$($syst.$)
^{+0.95}_{-1.96}$(pol.)$\mu$ and 
$\sigma_{\rm J/\psi}$($|y|<$0.9)=10.7$\pm$1.2(stat.)$\pm$1.7$($syst.$)
^{+1.6}_{-2.3}$(pol.)$\mu$b. 
Systematic uncertainties include the errors on the signal 
extraction, on the trigger and reconstruction efficiencies, on the 
$p_{\rm T}$ and $y$ input distributions used in the 
$A \times \epsilon$ evaluation, and on the luminosity 
determination.
The influence of the unknown degree of \jpsi\ polarization on the acceptance
evaluation is quoted separately.
Following the same approach adopted for the evalution of the integrated cross section, 
also the differential distributions are studied. 
The ALICE muon measurement and the LHCb result, covering the same $y$ range, are 
in good agreement, while the ALICE electron 
result complements the high $p_{\rm T}$ ATLAS and CMS distributions, reaching 
$p_{\rm T}$=0.
More details on the analysis can be found in~\cite{Aam11}.
 
\begin{figure}[htbp]
\vskip -0.5cm
\centering
\includegraphics[width=.47\textwidth]{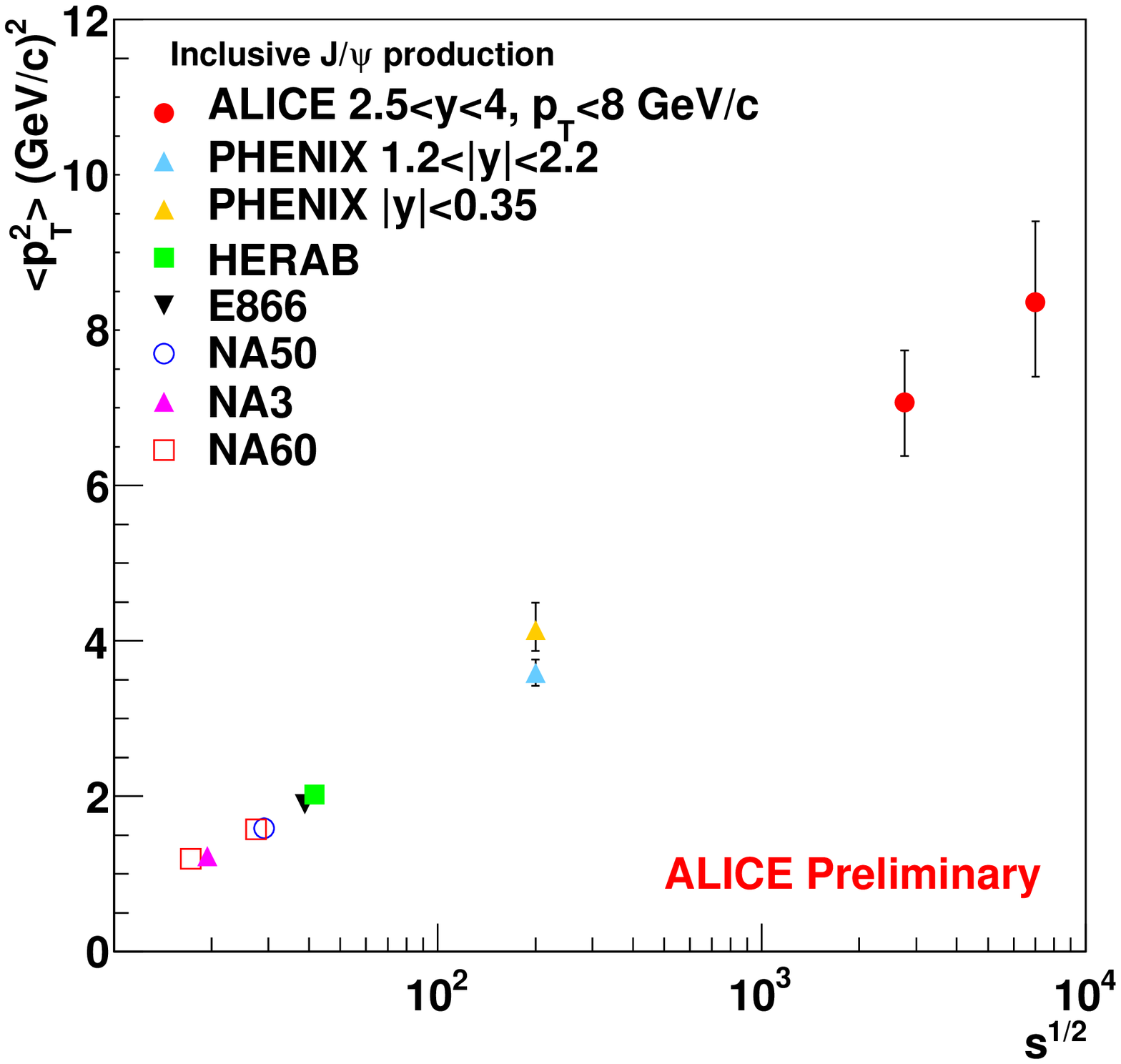}
\includegraphics[width=.425\textwidth]{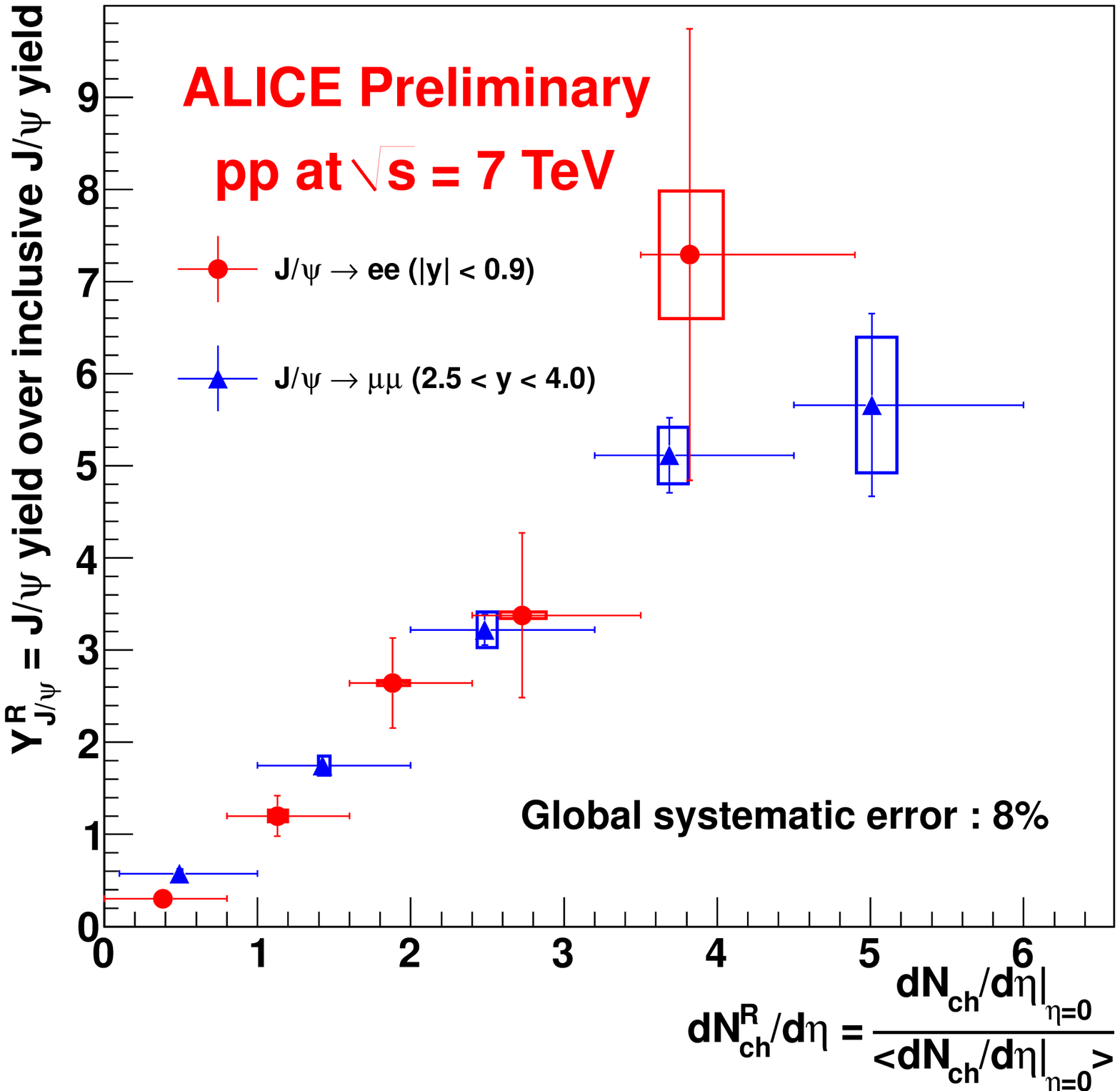}
\caption{Left: \jpsi\ $\langle p_{\rm T}^{2} \rangle$ versus 
$\sqrt{s}$. Right: $Y^{R}_{\rm J/\psi}$ versus $dN_{ch}^{R}/d\eta$}
\label{fig:2}
\vskip -0.4cm
\end{figure}

The analysis of the \jpsi\ production in pp collisions at $\sqrt{s}$=2.76~TeV, in the 
dimuon channel, is based on $L_{\rm int}=20.2$ nb$^{-1}$, 
while in the dielectron channel on $L_{\rm int}=1.1$ nb$^{-1}$. The
analysis technique is the same adopted for the data at $\sqrt{s}$=7~TeV, as
described above.
At $\sqrt{s}$=2.76~TeV, the number of \jpsi\ in the dimuon channel is 
$N_{\rm J/\psi}=1287\pm48$, while in the dielectron channel, 
$N_{\rm J/\psi}=41\pm9$. $A
\times \epsilon$ is 35$\%$ for the dimuon analysis and 9$\%$ for the
dielectron one.
The preliminary inclusive \jpsi\ production cross sections at $\sqrt{s}$=2.76~TeV are: 
$\sigma_{\rm J/\psi}(2.5<y<4)$=3.46$\pm$0.13(stat.)$\pm$0.42$($syst.$)^{+0.55}_{-1.11}$(pol.)$\mu$b and
$\sigma_{\rm J/\psi}(|y|<0.9)$=6.44$\pm$1.42(stat.)$\pm$1.03$($syst.$)^{+0.64}_{-1.42}$(pol.)$\mu$b. 
The sources and the magnitudes of systematics uncertainties are similar to those 
evaluated at $\sqrt{s}$=7~TeV.
These results, obtained at the same
center-of-mass energy of the Pb-Pb collisions, are mandatory for the evaluation
of the Pb-Pb nuclear modification factor $R_{AA}$, to correctly quantify the hot nuclear
matter effects on the \jpsi\ yield~\cite{Pil11}.
\begin{figure}[htbp]
\vskip -0.5cm
\centering
\includegraphics[width=.48\textwidth]{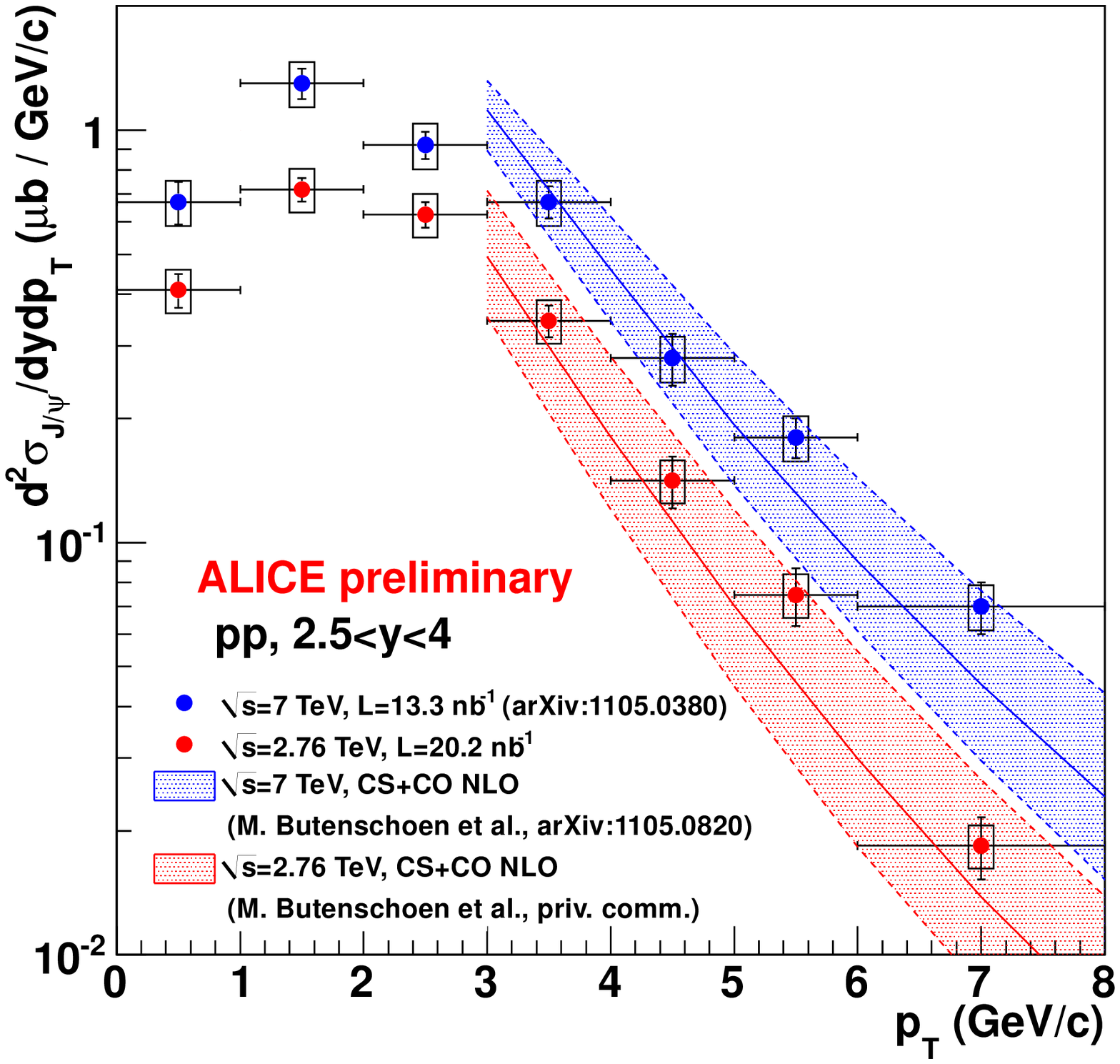}
\includegraphics[width=.48\textwidth]{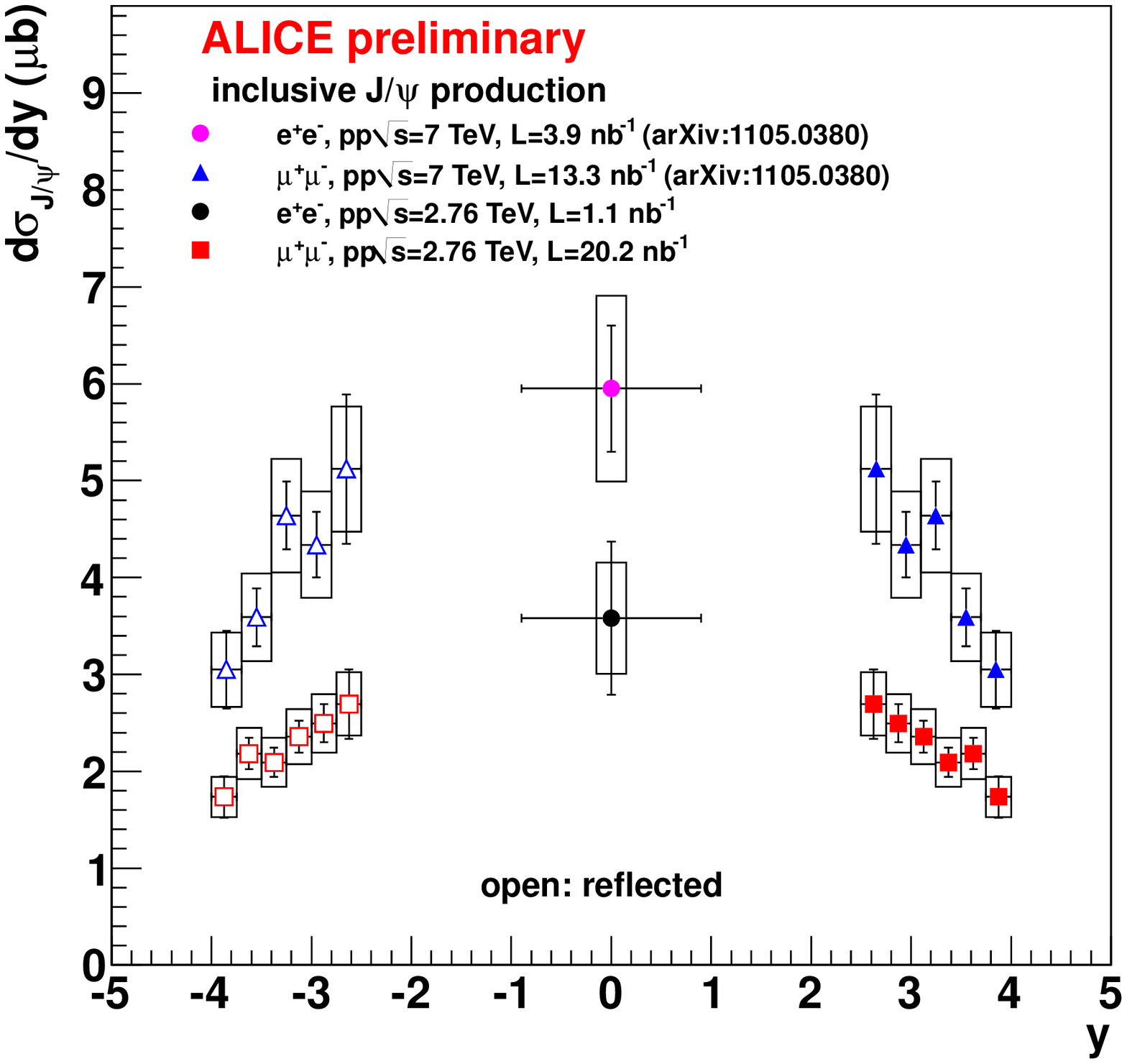}
\caption{Left: $d^{2}\sigma_{\rm J/\psi}/dydp_{\rm T}$ obtained at
$\sqrt{s}$=7~TeV and 2.76~TeV, compared with NLO NRQCD calculations. 
Right: $d\sigma_{\rm J/\psi}/dy$ obtained at $\sqrt{s}$=7~TeV and 2.76~TeV.}
\label{fig:1}
\end{figure}
The larger \jpsi\ statistics collected in the dimuon channel allows 
the evaluation of the differential $p_{\rm T}$ and $y$ cross 
sections. In Fig.~\ref{fig:1} the 
distributions are compared to those obtained at $\sqrt{s}$=7~TeV. At both
energies, NRQCD NLO calculations~\cite{But11}, available for $p_{\rm T}>3$~GeV/c, 
provide a good description of the $p_{\rm T}$ distributions.
The $d\sigma_{\rm J/\psi}/dy$ measured at 
midrapidity increases linearly as a function of $\sqrt{s}$, following the 
trend defined by RHIC and Tevatron results at lower energies.
Furthermore, the integrated cross section at both energies are in
reasonable agreement with expectations based on calculations from FONLL and leading order color 
evaporation model~\cite{Bos11}. 
Fitting the ALICE $p_{\rm T}$ differential distributions with the function 
$A\times p_{T}/(1+(p_{T}/p_{0})^2)^n$, the \jpsi\ 
$\langle p_{\rm T} \rangle$ and $\langle p_{\rm T}^{2} \rangle$ have been 
extracted.
A logarithmic increase of $\langle p_{\rm T}^{2} \rangle$ with $\sqrt{s}$ 
holds from low energy up to the LHC domain, as shown in Fig.~\ref{fig:2} (left),  and a similar trend is observed also for $\langle p_{\rm T}
\rangle$.
The \jpsi\ production yield at $\sqrt{s}$=7~TeV has also been studied as a 
function of the charged particle pseudo-rapidity density $dN_{ch}/d\eta$, measured in the ITS ($|\eta|<$1.6), both in the
dielectron and in the dimuon channel. The multiplicity corresponding to five
times the average value is of about 30, and it is similar to 
the one measured in semi-peripheral Cu-Cu interactions at $\sqrt{s_{NN}}$=200~GeV~\cite{Alv11}. Therefore, high multiplicity
pp collisions might present features observed, up to now, only in nucleus-nucleus
interactions at lower energy~\cite{Por10}. Following an approach similar to the one described
for the cross section evaluation, the \jpsi\ yield is extracted in intervals of 
$dN_{ch}/d\eta$. The relative yield $Y^{R}_{\rm J/\psi}$, i.e. the \jpsi\ yield in each
bin divided by the inclusive yield in minimum bias events, is shown in Fig.~\ref{fig:2} (right), as a
function of the relative charged particle density
($dN_{ch}^{R}/d\eta=dN_{ch}/d\eta|_{\eta=0}/\langle
dN_{ch}/d\eta|_{\eta=0}\rangle$).
Preliminary results in both the dielectron and dimuon channel show a similar linear increase as a
function of multiplicity. 

The ALICE experiment has measured the inclusive \jpsi\  
production at $\sqrt{s}$=2.76~TeV and 7~TeV in two rapidity ranges, $2.5<y<4$ 
and $|y|<0.9$. At both central and
forward rapidities the \jpsi\ production is measured down to $p_{\rm T}$=0, 
providing a unique acceptance coverage at LHC.
The collected pp data allow also a first analysis of \jpsi\ 
production as a function of particle multiplicity, and a linear increase of the
production yield is observed.
The analysis of the full statistics collected in 2010, together with the new
data from the on-going run, will provide further insights on the \jpsi\ production, 
addressing other topics, such as the study of the \jpsi\ polarization and the identification, through
the study of the pseudo-proper decay time distributions, of the
\jpsi\ coming from B decay.

\end{document}